\documentclass[aps,letterpaper,twocolumn,prl,showpacs]{revtex4-1}
\usepackage{amssymb}
\usepackage{amsmath,bm}
\usepackage{mathrsfs}

\usepackage[pdftex]{graphicx,color}
\usepackage[english]{babel}
\usepackage{hyperref}
\usepackage{epstopdf}
\begin{document}

\title{Experimental observation of internally-pumped parametric oscillation and quadratic comb generation in a $\chi^{(2)}$ whispering-gallery-mode microresonator}

\author{Ian Hendry$^{1,2}$}
\email{i.hendry@auckland.ac.nz}
\author{Luke S. Trainor$^{2,3}$}
\author{Yiqing Xu$^{1,2}$}
\author{St\'ephane Coen$^{1,2}$}
\author{Stuart G. Murdoch$^{1,2}$}
\author{Harald G. L. Schwefel$^{2,3}$}
\author{Miro Erkintalo$^{1,2}$}
\email{m.erkintalo@auckland.ac.nz}

\affiliation{$^1$The Department of Physics, The University of Auckland, Auckland 1010, New Zealand}
\affiliation{$^2$The Dodd-Walls Centre for Photonic and Quantum Technologies, New Zealand}
\affiliation{$^3$Department of Physics, University of Otago, Dunedin 9016, New Zealand}

\begin{abstract}
We report on the experimental observation of internally-pumped parametric oscillation in a high-Q lithium niobate microresonator under conditions of natural phase-matching. Specifically, launching near-infrared pump light around 1060~nm into a $z$-cut congruent lithium niobate microresonator, we observe the generation of optical sidebands around the input pump under conditions where second-harmonic generation is close to natural phase-matching. We find that a wide range of different sideband frequency shifts can be generated by varying the experimental parameters. Under particular conditions, we observe the cascaded generation of several equally-spaced sidebands around the pump -- the first steps of optical frequency comb generation via cavity-enhanced second-harmonic generation.
\end{abstract}

\maketitle

\noindent High-Q microresonators enable strong nonlinear optical effects to be driven with low-power continuous wave laser light, allowing for novel nonlinear optical devices~\cite{strekalov_nonlinear_2016, breunig_three-wave_2016, meisenheimer_continuous-wave_2017, sayson_octave-spanning_2019}. The generation of optical frequency combs in third-order, Kerr nonlinear microresonators has attracted particular attention~\cite{kippenberg_microresonator-based_2011, pasquazi_micro-combs:_2018, gaeta_photonic-chip-based_2019}, with demonstrated applications ranging from telecommunications~\cite{marin-palomo_microresonator-based_2017} to distance measurements~\cite{suh_soliton_2018}. Interestingly, recent studies have shown that it is possible to generate frequency combs also via second-order $\chi^{(2)}$ nonlinear processes; experimental observations have been reported in resonators designed for second-harmonic generation (SHG)~\cite{ulvila_frequency_2013,ricciardi_frequency_2015}, parametric down-conversion (PDC)~\cite{mosca_modulation_2018}, and the electro-optic effect~\cite{rueda_resonant_2019, zhang_broadband_2019}. Such \emph{quadratic} combs are attracting increasing attention due to their potential advantages compared to their third-order Kerr counterparts.

\begin{figure}[t]	
		\includegraphics[width=\linewidth]{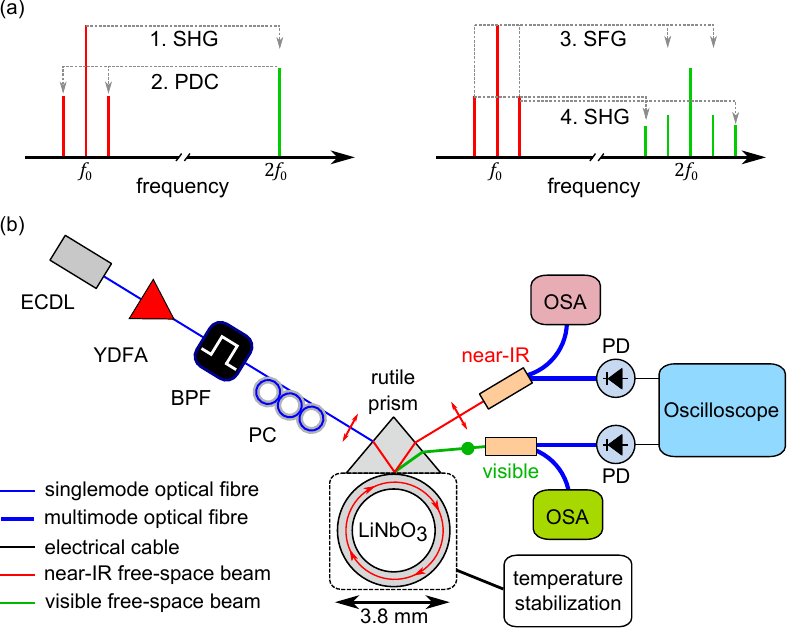}
		\caption{(a) Schematic illustration of the generation of spectral sidebands around the fundamental and second-harmonic frequencies $f_0$ and $2f_0$, respectively. SFG, sum-frequency generation. (b) Schematic illustration of the experimental setup. ECDL, exernal-cavity diode laser; YDFA, ytterbium-doped fibre amplifier; BPF, band-pass filter; PC, polarization controller; OSA, optical spectrum analyzer; PD, photodetector.}
		\label{fig1}
\end{figure}

Cavity-enhanced SHG is attractive for offering a route to generate frequency combs in the visible using a near-infrared pump. The first experimental demonstrations of comb generation via SHG were reported in a macroscopic free-space system under conditions of large phase mismatch~\cite{ulvila_frequency_2013} -- a regime well-known to \emph{mimic} the $\chi^{(3)}$ Kerr nonlinearity~\cite{stegeman_2_1996}. Later studies~\cite{ricciardi_frequency_2015} unveiled that frequency combs can be generated even under conditions of perfect phase-matching through so-called internally-pumped optical parametric oscillation (OPO)~\cite{marte_competing_1994, schiller_subharmonicpumped_1996}: the second-harmonic signal acts as an internally-generated pump for a PDC process that produces equally-spaced spectral sidebands around the pump [see Fig.~\ref{fig1}(a)]. Whilst the initial demonstrations of comb generation in cavity-enhanced SHG were achieved using macroscopic free-space resonators~\cite{ricciardi_frequency_2015}, recent experiments performed in periodically-poled waveguide Fabry-Perot resonators have shown strong evidence that the concept can be translated into a miniaturised format~\cite{ikuta_frequency_2018, stefszky_towards_2018}.

Monolithic whispering-gallery mode (WGM) microresonators can display unprecedented Q-factors compared to conventional resonators, offering significant potential benefits for comb generation via cavity-enhanced SHG. Yet, whilst impressive results pertaining to SHG and PDC have been achieved in such systems~\cite{meisenheimer_continuous-wave_2017, ilchenko_nonlinear_2004, furst_naturally_2010, beckmann_highly_2011, schunk_interfacing_2015, fortsch_highly_2015}, signatures of comb generation have remained elusive. In this Letter, we report on experimental observations of internally-pumped OPO and the initial stages of comb generation in a lithium niobate WGM resonator. We pump the resonator with near-infrared (near-IR) light at around 1060~nm, and use thermal tuning to naturally phase-match SHG. Under suitable conditions, we observe the generation of equally-spaced sidebands around the pump -- a characteristic feature of internally-pumped OPO. Moreover, for particular pump wavelengths and resonator temperatures, we observe the initial stages of a cascade that results in the generation of several sideband orders. Our results represent a proof-of-concept demonstration of SHG comb generation in naturally phase-matched WGM microresonators, and could help realise novel sources of optical frequency combs.

Our experimental configuration, which is schematically illustrated in Fig.~\ref{fig1}(b), is similar to the one used previously for observations of naturally-phase matched SHG in WGM resonators~\cite{furst_naturally_2010}. The resonator used was fabricated from a $z$-cut, 5\% MgO-doped congruent lithium niobate window via single-point diamond turning and subsequent mechanical polishing. It has a radius of 1.9~mm, corresponding to a free-spectral range of about 11.2~GHz. We measured a typical resonance width to be about 5~MHz around 1060~nm, corresponding to a Q-factor of $5.6\times10^7$ and a finesse of 2300.

Our driving source is a fibre-coupled external cavity diode laser that is tunable from 1030~nm to 1070~nm. An ytterbium-doped fibre amplifier is used to further amplify the laser output, and a tunable bandpass filter is used to remove amplified spontaneous emission. The pump is coupled into the resonator using frustrated total internal reflection from a rutile prism; a pigtailed ferrule and a GRIN lens is used to focus the pump on the back-side of the prism, whilst piezo-stages allow for fine-tuning of the distance between the prism and the resonator.  At the output of the prism, we detect the signals around the pump and its second-harmonic using two photodetectors connected to an oscilloscope as well as two optical spectrum analyzers (OSAs) that operate in the near-IR and the visible, respectively.

To achieve phase-matching for SHG, the pump is coupled into an ordinary polarized mode (polarization perpendicular to the optical axis), whilst the second-harmonic is generated in an extra-ordinary mode (polarization along the optical axis).  The resonator is mounted on a brass rod whose temperature is stabilized to a set level so as to thermally tune the phase-matching condition. By testing several different temperatures, we found that $94^\circ\mathrm{C}$ yielded optimal SHG. This value is close to the phase-matching temperature found in~\cite{furst_naturally_2010}. Additional, fine thermal control is achieved by shining a blue laser with variable intensity on the resonator mode volume.

\begin{figure}[t]	
		\includegraphics[width=\linewidth]{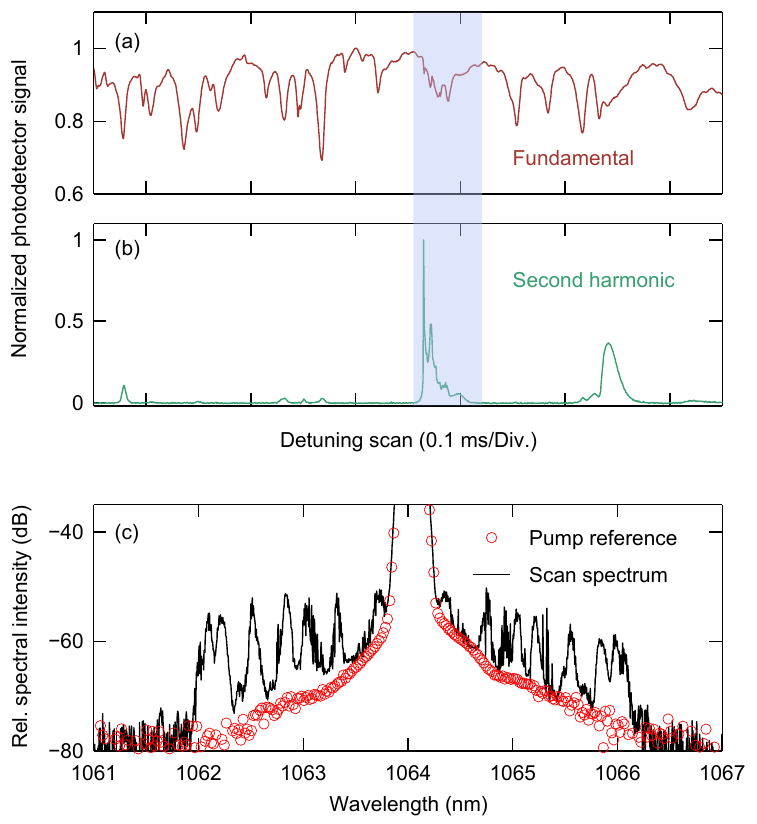}
		\caption{(a) Fundamental and (b) second harmonic intensity measured at the resonator output as the laser frequency is linearly scanned in time. Both curves are normalized to unity. (c) Solid black curve shows the optical spectrum measured around the fundamental wavelength at the resonator output when the laser frequency is scanned over a single cavity resonance indicated by the shaded area in (a) and (b). The open red circles depict the pump spectrum for reference.}
		\label{fig2}
\end{figure}

\looseness=-1 With the temperature stabilized at $94^\circ\mathrm{C}$, we find that several different mode families produce observable SHG. Figures~\ref{fig2}(a) and (b) show illustrative oscilloscope recordings of the fundamental and second-harmonic intensities, respectively, when the laser frequency is scanned over several modes. Here the fundamental power right before the pigtailed ferrule was measured to be about 100~mW. At this power level, we find that the SHG produced by some of the resonator modes is sufficiently strong so as to act as an internally-generated pump for an OPO process, giving rise to spectral sidebands around the pump. Indeed, Fig.~\ref{fig2}(c) shows a spectral measurement around the pump, obtained as the pump frequency is scanned over one of the cavity resonances [indicated by the shaded area in Figs.~\ref{fig2}(a) and (b)]. The broadening of the pump spectrum is obvious. Note that the OSA sweep speed is much slower than the pump frequency scan, and so the spectrum in Fig.~\ref{fig2}(c) traces out the envelope of all the different signals that can be generated at different detunings. Moreover, the resolution of our OSA is larger than the 11.2~GHz free-spectral range of the resonator, prohibiting us from distinguishing the presence of individual lines with that spacing.

\begin{figure}[t]	
		\includegraphics[width=1\linewidth]{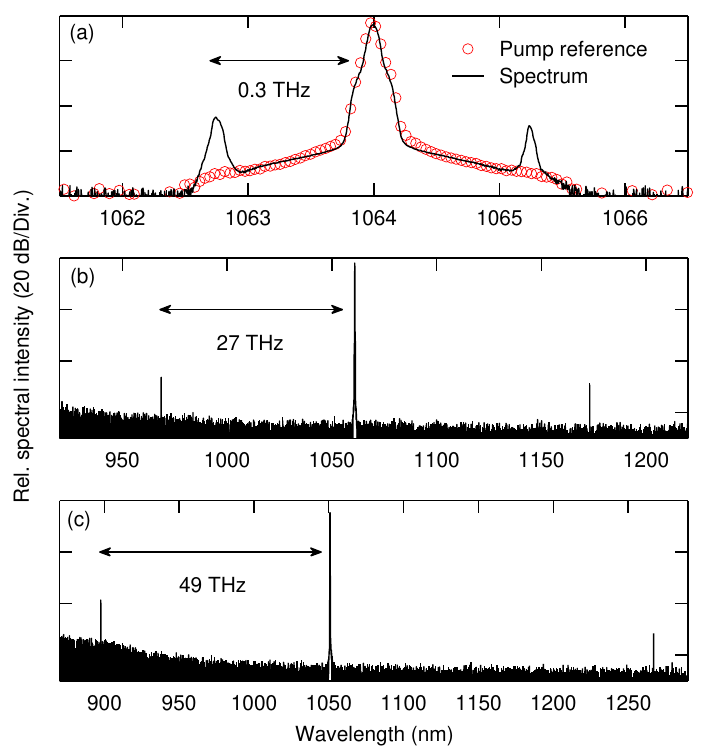}
		\caption{Experimentally measured spectra around the pump obtained for different pump wavelengths: (a) 1064~nm, \mbox{(b) 1061~nm}, (c) 1051~nm. The open red circles in (a) indicate the pump spectrum for reference.}
		\label{fig3}
\end{figure}

To observe steady-state signals, we tune the pump laser into selected resonances from the blue, leveraging passive thermal locking to subsequently maintain fixed detuning~\cite{carmon_dynamical_2004}. For each resonance, we fine-adjust the coupling and resonator temperature so as to optimise nonlinear frequency conversion. Figures~\ref{fig3}(a)--(c) show illustrative examples of spectra measured around the pump for different pump wavelengths. We clearly see optical sidebands generated via the PDC of the internally-generated SHG signal. By pumping different WGM resonances, we can observe a wide variety of different sideband frequency shifts, ranging from hundreds of GHz [Fig.~\ref{fig3}(a)] to almost 50~THz [Fig.~\ref{fig3}(c)]. Even when pumping a single resonance, we find that small adjustments of the pump-resonator detuning can yield sidebands with noticeably different frequency shifts. Detailed mapping of the entire tuning characteristics is, however, beyond the scope of our present Letter.

\begin{figure}[t]	
		\includegraphics[width=1\linewidth]{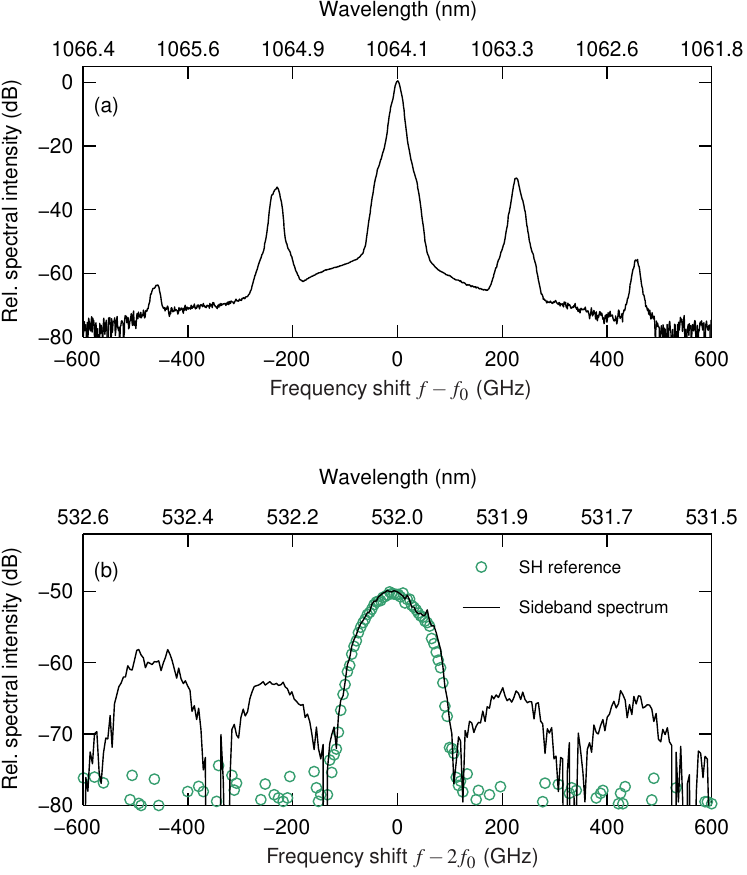}
		\caption{Experimentally measured spectrum around the (a) pump and (b) second-harmonic wavelengths. For reference, the green open circles in (b) show the second-harmonic (SH) spectrum in the absence of sidebands around the fundamental, normalized such that its noise floor is similar to the main recording shown as the black solid curve.}
		\label{fig4}
\end{figure}

At certain pump wavelengths, we observe the cascaded generation of several sideband pairs. Figures~\ref{fig4}(a) and (b) show illustrative spectra measured around the pump and the second-harmonic, respectively, when operating under such conditions. We clearly see the emergence of second-order sidebands around the fundamental, representing the initial stages of quadratic comb generation~\cite{ricciardi_frequency_2015}. New frequency components are also generated around the second-harmonic [Fig.~\ref{fig4}(b)] at wavelengths corresponding to sum-frequency generation and SHG processes involving the spectral components around the fundamental. We must note that the resolution of our spectral measurements around the second-harmonic is limited by the OSA used in that frequency range.

\begin{figure}[t]	
		\includegraphics[width=\linewidth]{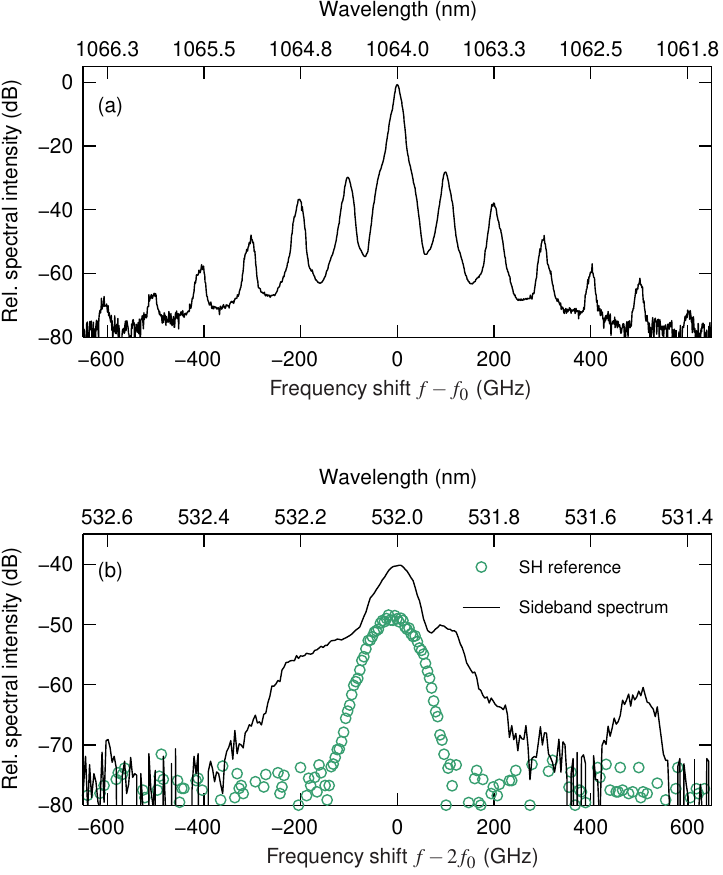}
		\caption{Experimentally measured spectrum around the (a) pump and (b) second-harmonic wavelengths. For reference, the green open circles in (b) show the second-harmonic (SH) spectrum in the absence of sidebands around the fundamental, normalized such that its noise floor is similar to the main recording shown as the black solid curve.}
		\label{fig5}
\end{figure}

\looseness=-1Figure~\ref{fig5} shows an additional example of the initial stages of quadratic comb generation. Here we see six pairs of spectral sidebands around the pump [Fig.~\ref{fig5}(a)], spaced from one another by about 100~GHz in frequency (corresponding to nine free-spectral ranges). Because of the smaller sideband spacing compared to results shown in Fig.~\ref{fig4}, we are not able to resolve the individual sidebands around the second-harmonic; rather, their presence is manifest in the broadening of the second-harmonic spectrum [Fig.~\ref{fig5}(b)].

Quantitative comparison between our experimental findings and theoretical results developed in e.g.~\cite{leo_walk-off-induced_2016,leo_frequency-comb_2016, hansson_single_2016, villois_soliton_2019} is prohibited by poor knowledge over salient model parameters. In particular, our current experimental configuration does not allow us to identify the precise mode families involved in the nonlinear interactions, which results in significant uncertainty in the nonlinear coupling coefficients. Moreover, we do not currently have the capability to measure the value of the pump-cavity detuning nor the residual phase-mismatch, both of which can significantly influence the nonlinear cavity dynamics. We must nevertheless emphasize that the experimental observations reported above cannot be ascribed to direct $\chi^{(3)}$ Kerr nonlinearity. Indeed, lithium niobate exhibits very large normal dispersion around 1060~nm, which is prohibitive for the phase-matching of degenerate four-wave mixing. The fact that our observations of sideband generation is always correlated with strong SHG provides further evidence of the sidebands' $\chi^{(2)}$ origins.

While our present experimental configuration permits us to reliably generate sidebands around the pump and the second-harmonic, we are unable to maintain stable operation for more than some tens of seconds at a time. In addition to sub-optimal thermal isolation, the linewidth of our pump laser ($\sim 200$~kHz) necessitates sacrifices with regards to the resonator Q-factor, which in turn leads to increased pump power requirements that can impede system stability. While finalising our manuscript, encouraging results with 2~mW comb generation power threshold was reported in a system otherwise similar to ours but with a better resonator Q-factor and pump laser linewidth~\cite{szabados_frequency_2019}. Finally, we must note that we have observed in our system microsecond-scale self-pulsing instabilities, whose origins we speculate to lie in thermal and/or photorefractive effects~\cite{luo_MHz-level_2014, sun_nonlinear_2017}. Such instabilities can occur simultaneously with sideband generation, highlighting the need to carefully assess the time-domain system stability during sideband or comb generation. We have taken particular care to ensure all the results reported above were obtained in the absence of self-pulsing effects.

To conclude, we have reported on the experimental observation of internally-pumped parametric oscillation and initial stages of quadratic comb formation in a $\chi^{(2)}$ WGM microresonator. Whilst our current experimental configuration has enabled the proof-of-principle demonstrations reported in this work, there are a number of improvements that are left for future work. For example, operation in the fundamental mode family is envisaged to significantly enhance the nonlinear interactions, whilst better temperature control is expected to improve the phase-matching as well as the overall system stability. Our work nevertheless demonstrates the viability of observing and studying cascaded quadratic nonlinear phenomena in WGM microresonators.

\section*{Acknowledgements}
\noindent We acknowledge financial support from the Marsden Fund and the Rutherford Discovery Fellowships of the Royal Society of New Zealand. We also acknowledge useful discussions with Ingo Breunig and Jan Szabados.

\end{document}